\documentclass{llncs}
\usepackage{psfig}
\begin{document}
\frontmatter          
\pagestyle{headings}  
\addtocmark{Path integral Monte Carlo simulations} 

\title{Path integral Monte Carlo simulations  and analytical approximations for
high-temperature plasmas}
\author{
V. Filinov\inst{1,2}, M. Bonitz\inst{1}, D. Kremp\inst{1},
W.-D.Kraeft\inst{3}, and V.Fortov\inst{2}
}
\institute{Fachbereich Physik, Universit{\"a}t Rostock\\
Universit{\"a}tsplatz 3, D-18051 Rostock, Germany \\
\and
Institute for High Energy Density, Russian Academy of Sciences,  \\
ul. Izhorskaya 13/19, Moscow, 127412 Russia \\
E - mail: vs\_filinov@hotmail.com \\
\and
Institut f{\"u}r Physik, Universit{\"a}t Greifswald\\
Domstr.10a, D-17487 Greifswald, Germany
}
\maketitle

%
\makeatletter
\renewenvironment{thebibliography}[1]
     {\section*{\refname}
      \small
      \list{}%
           {\settowidth\labelwidth{}%
            \leftmargin\parindent
            \itemindent=-\parindent
            \labelsep=\z@
            \if@openbib
              \advance\leftmargin\bibindent
              \itemindent -\bibindent
              \listparindent \itemindent
              \parsep \z@
            \fi
            \usecounter{enumiv}%
            \let\p@enumiv\@empty
            \renewcommand\theenumiv{}}%
      \if@openbib
        \renewcommand\newblock{\par}%
      \else
        \renewcommand\newblock{\hskip .11em \@plus.33em \@minus.07em}%
      \fi
      \sloppy\clubpenalty4000\widowpenalty4000%
      \sfcode`\.=\@m}
     {\def\@noitemerr
       {\@latex@warning{Empty `thebibliography' environment}}%
      \endlist}
      \def\@cite#1{#1}%
      \def\@lbibitem[#1]#2{\item[]\if@filesw
        {\def\protect##1{\string ##1\space}\immediate
      \write\@auxout{\string\bibcite{#2}{#1}}}\fi\ignorespaces}
\makeatother
%
\begin{abstract} The results of analytical approximations and extensive
calculations based on a path integral Monte Carlo (PIMC) scheme
 are presented. A new (direct) PIMC method allows for a correct
determination of thermodynamic properties such as energy and equation
of state of dense degenerate Coulomb systems. In this paper, we present
results for dense partially ionized hydrogen at intermediate and high temperature.
We give a  quantitative comparison with the available results
of alternative (restricted) PIMC simulations and with analytical
expressions based on iterpolation formulas meeting the exact limits at low
and high densities. Good agreement between the two simulations is found
up to densities of the order of $10^{24}cm^{-3}$. The agreement with the
analytical results is satisfactory up to densities in the range
$10^{22}\dots 10^{23}cm^{-3}$.
\end{abstract}
\section{Introduction} Correlated Fermi systems are of increasing interest in
many fields, including plasmas, astrophysics, solids and nuclear
matter, (see
Kraeft et al. \cite{green-book}) for an overview. Among
the topics of
current interest are Fermi liquids, metallic hydrogen (see
DaSilva et al. \cite{dasilva-etal.97}),
plasma phase transition (see Schlanges et al. \cite{Schla95}), bound states etc.
In such many particle quantum systems, the Coulomb interaction is
essential. There has been significant
progress in recent years to study these systems theoretically,
and especially numerically, (see e.g.
Bonitz (Ed.) \cite{kbt99}, Zamalin et al. \cite{zamalin},
Filinov, A. V. et
al. \cite{afilinov-etal.00pss}).
A theoretical framework which is
particularly well suited to describe thermodynamic  properties in the region of
strong coupling and degeneracy is the path integral quantum Monte Carlo (PIMC)
method. There has been  remarkable recent progress in applying these techniques
to  Fermi systems.
However, these
simulations are essentially hampered by the fermion sign  problem.  To overcome
this difficulty, several strategies have been developed to  simulate macroscopic
Coulomb systems
(see Militzer and Pollock \cite{mil-pol}, Militzer and Ceperley
 \cite{militzer-etal.00}, and Militzer \cite{militzer_phd}):
the  first is the restricted PIMC concept where additional assumptions on the
density operator
${\hat \rho}$ are introduced which reduce the sum over  permutations to even
(positive) contributions only. This requires the knowledge  of the nodes of the
density matrix which is available only in a few special cases. However,
for interacting  macroscopic systems, these nodes are known only approximately,
(see, e.g.,  Militzer and Pollock \cite{mil-pol} and Militzer and
Ceperley \cite{militzer-etal.00}), and the accuracy
of the results is difficult to assess from within this scheme.

Recently, we have published a new path
integral representation  for the N-particle density operator
(see Filinov, V. S.,  et al. \cite{filinov-etal.00jetpl}, Bonitz
(Ed.) \cite{kbt99},
Filinov, V. S., et al. \cite{ebeling-etal.00}),
which  allows for {\em direct
Fermionic path integral Monte Carlo} simulations  of dense plasmas in a broad
range of densities and temperatures. Using this  concept we computed the
pressure (equation of state, EOS), the energy,
and the  pair distribution functions
of a dense partially ionized and dissociated electron--proten
plasma (see Filinov, V. S., et al. \cite{filinov-etal.00jetpl}).
In this region no reliable
data are available from
other theories such as density  functional theory or quantum statistics
(see, e.g., Kraeft et al. \cite{green-book}), which would allow for an
unambiguous test.
Therefore, it is of high interest  to perform quantitative comparisons of
analytical results and
independent numerical simulations, such as  restricted and direct fermionic
PIMC, which is the aim of this paper.

\section{Path integral representation of thermodynamic quantities}
We now briefly outline the idea of our direct PIMC scheme.
All thermodynamic properties of a two-component plasma are defined
by the partition function $Z$ which, for the case of $N_e$
electrons and $N_p$ protons, is given by
\begin{eqnarray}
Z(N_e,N_p,V,\beta) &=&
\frac{Q(N_e,N_p,\beta)}{N_e!N_p!},
\nonumber\\
\mbox{with} \qquad
Q(N_e,N_p,\beta) &=& \sum_{\sigma}\int\limits_V dq \,dr
\,\rho(q, r,\sigma;\beta),
\label{q-def}
\end{eqnarray}
where $\beta=1/k_B T$.
The exact density matrix is, for a quantum system, in
general, not known but can be constructed using a path integral representation
(see Feynman and Hibbs \cite{feynman-hibbs}),
\begin{eqnarray}
\int\limits_{V} dR^{(0)}\sum_{\sigma}\rho(R^{(0)},\sigma;\beta) &=&
 \int\limits_{V} dR^{(0)} \dots dR^{(n)} \,
\rho^{(1)}\cdot\rho^{(2)} \, \dots \rho^{(n)}
\nonumber\\
&\times&\sum_{\sigma}\sum_{P} (\pm 1)^{\kappa_P}
\,{\cal S}(\sigma, {\hat P} \sigma')\,
{\hat P} \rho^{(n+1)},
\label{rho-pimc}
\end{eqnarray}
where $\rho^{(i)}\equiv \rho\left(R^{(i-1)},R^{(i)};\Delta\beta\right) \equiv
\langle R^{(i-1)}|e^{-\Delta \beta {\hat H}}|R^{(i)}\rangle$,
whereas $\Delta \beta \equiv \beta/(n+1)$.
${\hat H}$ is the Hamilton operator, ${\hat H}={\hat K}+{\hat U}_c$, containing
kinetic and potential energy contributions with
${\hat  U}_c = {\hat  U}_c^p + {\hat  U}_c^e + {\hat  U}_c^{ep}$ being the sum
of the Coulomb potentials between protons (p), electrons (e) and electrons and
protons (ep).
Further,
$R^{(i)}=(q^{(i)},r^{(i)}) \equiv (R_p^{(i)},R_e^{(i)})$, for $i=1,\dots n+1$,
$R^{(0)}\equiv (q,r)\equiv (R_p^{(0)},R_e^{(0)})$. Also,
$R^{(n+1)} \equiv R^{(0)}$ and $\sigma'=\sigma$, i.e., the particles are
represented by closed Fermionic loops with the coordinates (beads)
$[R]\equiv [R^{(0)}; R^{(1)};\dots; R^{(n)}; R^{(n+1)}]$, where $r$ and
$q$ denote the electron and proton coordinates, respectively.
The spin gives rise to the spin part of the density matrix ${\cal S}$, whereas
exchange effects are accounted for by the permutation operator ${\hat P}$, which
acts on the electron coordinates and spin, and
the sum over the permutations with parity $\kappa_P$. In the fermionic case
(minus sign), the sum contains $N_e!/2$ positive and negative terms leading
to the notorious sign problem. Due to the large mass difference of electrons
and protons, the exchange of the latter is not included.

Recently, we have derived a new representation for the
high--temperature density matrices $\rho^{(i)}$ in eq.(2) (see
Filinov, V. S., et al. \cite{filinov-etal.00jetpl}) which is well suited for direct PIMC
simulations. A crucial point is that the electron--proton interaction can be
described by an (effective) quantum pair potential $\Phi^{ep}$ ({\it
Kelbg--potential}, see Kelbg \cite{Ke63}). For details of the
derivation see Filinov, V. S., et al. \cite{filinov-etal.00jetpl}. Here, we present only the
final result for the energy and for the EOS.
Consider first the energy:
\begin{eqnarray}
\beta E = \frac{3}{2}(N_e+N_p) + \frac{1}{Q}
\frac{1}{\,\lambda_p^{3N_p}\Delta \lambda_e^{3N_e}}\sum_{s=0}^{N_e}
\int dq \, dr \, d\xi \,\rho_s(q,[r],\beta) \,\times
\nonumber\\
\Bigg\{\sum_{p<t}^{N_p} \frac{\beta e^2}{|q_{pt}|} +
\sum_{l=0}^{n}\Bigg[\sum_{p<t}^{N_e} \frac{\Delta\beta e^2}{|r^l_{pt}|}
+  \sum_{p=1}^{N_p}\sum_{t=1}^{N_e} \Psi_l^{ep}\Bigg]
\nonumber\\
+ \sum_{l=1}^{n}\Bigg[
- \sum_{p<t}^{N_e}C^l_{pt}
\frac{\Delta\beta e^2}{|r^l_{pt}|^2} +
 \sum_{p=1}^{N_p}\sum_{t=1}^{N_e}
D^l_{pt}
\frac{\partial \Delta\beta\Phi^{ep}}{\partial |x^l_{pt}|}
 \Bigg]
\nonumber\\
\,-\,
\frac{1}{{\rm det} |\psi^{n,1}_{ab}|_s}
\frac{\partial{\rm \,det} | \psi^{n,1}_{ab} |_s}{\partial \beta}
\Bigg\},
\nonumber \\
{\rm with} \quad C^l_{pt} = \frac{\langle
r^l_{pt}|y^l_{pt}\rangle}{2|r^l_{pt}|},
\qquad D^l_{pt} = \frac{\langle x^l_{pt}|y^l_{p}\rangle}{2|x^l_{pt}|},
\label{energy}
\quad
\end{eqnarray}
and $\Psi_l^{ep}\equiv \Delta\beta\partial
[\beta'\Phi^{ep}(|x^l_{pt}|,\beta')]/\partial\beta'|_{\beta'=\Delta\beta}$
contains the electron-proton Kelbg potential $\Phi^{ep}$.
Here,
$\langle \dots | \dots \rangle$ denotes the scalar product, and
$q_{pt}$, $r_{pt}$ and $x_{pt}$ are differences of two
coordinate vectors:
$q_{pt}\equiv q_p-q_t$,
$r_{pt}\equiv r_{p}-r_{t}$, $x_{pt}\equiv r_p-q_t$, $r^l_{pt}=r_{pt}+y_{pt}^l$,
 $x^l_{pt}\equiv x_{pt}+y^l_p$ and
 $y^l_{pt}\equiv y^l_{p}-y^l_{t}$, with
$y_a^n=\Delta\lambda_e\sum_{k=1}^{n}\xi^{(k)}_a$ and
$\Delta\lambda_a^2=2\pi\hbar^2 \Delta\beta/m_a$.
We introduced dimensionless distances between
neighboring vertices on the loop, $\xi^{(1)}, \dots \xi^{(n)}$,
thus, explicitly, $[r]\equiv [r; y_e^{(1)}; y_e^{(2)}; \dots]$.

The density matrix $\rho_s$ is given by
\begin{eqnarray}
\rho_s(q,[r],\beta) = C^s_{N_e}
\, e^{-\beta U(q,[r],\beta)} \prod\limits_{l=1}^n
\prod\limits_{p=1}^{N_e} \phi^l_{pp}
{\rm det} \,|\psi^{n,1}_{ab}|_s,
\label{rho_s}
\end{eqnarray}
where
$U(q,[r],\beta)=
U_c^p(q)+\{U^e([r],\Delta\beta)+U^{ep}(q,[r],\Delta\beta)\}/(n+1)$
and  $\phi^l_{pp}\equiv \exp[-\pi |\xi^{(l)}_p|^2]$.
We underline that the density matrix (\ref{rho_s})
does not contain an explicit
sum over the permutations and thus no sum of terms with alternating
sign. Instead, the whole exchange problem is
contained in a single exchange matrix given by
\begin{eqnarray}
||\psi^{n,1}_{ab}||_s\equiv ||e^{-\frac{\pi}{\Delta\lambda_e^2}
\left|(r_a-r_b)+ y_a^n\right|^2}||_s.
\label{psi}
\end{eqnarray}
As a result of the spin summation,
the matrix carries a subscript $s$ denoting the number of electrons having
the same spin projection. For more detail,
(see Filinov, V. S., et al. \cite{filinov-etal.00jetpl}, Bonitz
(Ed.) \cite{kbt99}).
In a similar way, we obtain the result for the equation of state,
\begin{eqnarray}
\frac{\beta p V}{N_e+N_p} = 1 + \frac{1}{N_e+N_p}
\frac{(3Q)^{-1}}{\,\lambda_p^{3N_p}\Delta\lambda_e^{3N_e}}\sum_{s=0}^{N_e}
\int dq \, dr \, d\xi \,\rho_s(q,[r],\beta) \times
\nonumber\\
\Bigg\{\sum_{p<t}^{N_p} \frac{\beta e^2}{|q_{pt}|} +
\sum_{p<t}^{N_e}\frac{\Delta\beta e^2}{|r_{pt}|}
-  \sum_{p=1}^{N_p}\sum_{t=1}^{N_e} |x_{pt}|
\frac{\partial \Delta\beta\Phi^{ep}}{\partial |x_{pt}|}
\nonumber\\
+\sum_{l=1}^{n}\left[\sum_{p<t}^{N_e}
A^l_{pt}
\frac{\Delta\beta e^2}{|r^l_{pt}|^2}
- \sum_{p=1}^{N_p}\sum_{t=1}^{N_e}B^l_{pt}
\frac{\partial \Delta\beta\Phi^{ep}}{\partial |x^l_{pt}|}
\right]
\nonumber\\
\,+\,\frac{\alpha}{{\rm det} |\psi^{n,1}_{ab}|_s}
\frac{\partial{\rm \,det} | \psi^{n,1}_{ab} |_s}{\partial \alpha}
\Bigg\},
\nonumber \\
{\rm with} \quad A^l_{pt} = \frac{\langle r^l_{pt}|r_{pt}\rangle}{|r^l_{pt}|},
\qquad B^l_{pt} = \frac{\langle x^l_{pt}|x_{pt}\rangle}{|x^l_{pt}|}.
\label{eos}
\quad
\end{eqnarray}

\section{Analytical approximations for the thermodynamic functions of dense
plasmas}

To describe dense plasmas, it is necessary to have thermodynamic functions valid at  arbitrary
degeneracy. Here, we restrict ourselves to the Hartree--Fock (HF)
and the Montroll--Ward (MW) contributions. This approximation is
appropriate at temperatures high enough such that the Coulomb
interaction is weak and the possibility of the
formation of bound states is excluded.
HF and MW contributions have been computed numerically (see Kraeft et
al. \cite{green-book}). The
analytical evaluation of the MW contribution is possible in  limiting situations
only, namely in the low and very high density  cases.  In the intermediate
region Pad{\'e} formulae can be used to interpolate  between the  limiting cases. In
between, the formulae are fitted to numerical data; (see
Ebeling et al. \cite{richert}
, Haronska et al. \cite{haro} and Ebeling and
Richert \cite{Pade85}).
We give the excess free energy and the
interaction part of  the chemical potential of the  electron gas,
\begin{equation}\label{pade1}
f^P=\frac{f_D-\frac{1}{4}(\pi\beta)^{-1/2}{\bar n}+8{\bar
n}^2f_{GB}}{1+8ln\left[1+\frac{3}{64\sqrt{2}}(\pi\beta)^{1/4}{\bar
n}^{1/2}\right]+8{\bar n}^2}\,,
\end{equation}
and
\begin{equation}\label{pade2}
\mu^P=\frac{\mu_D-
\frac{1}{2}(\pi\beta)^{-1/2}{\bar n}+8{\bar
n}^2\mu_{GB}}{1+8ln\left[1+\frac{1}{16\sqrt{2}}(\pi\beta)^{1/4}{\bar
n}^{1/2}\right]+8{\bar n}^2}\,.
\end{equation}
In (\ref{pade1},\ref{pade2}) Heaviside units  $\hbar=\frac{e^2}{2}=2m_e=1$ and
the  dimensionless  density
${\bar n}=n\Lambda^3$ were used.
\begin{figure}
\centerline{\psfig{figure=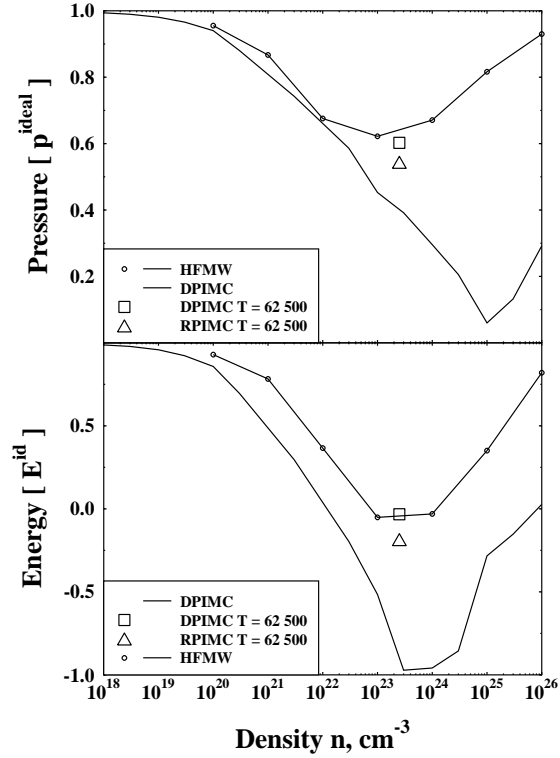,height=10cm}}
\caption{
Energy and pressure isotherms for $50,000 K$ (solid line).
Solid line with circles-- Hartree Fock (HF)  and Montroll--Ward
(MW) approximation.
Reference data: triangle -- RPIMC (see Militzer et al.), square -- DPIMC.}
\end{figure}

In  formulae  (\ref{pade1},\ref{pade2}), the
correct low density behaviour (Debye limiting law) is guaranteed
by choosing
$f_D=-(2/3)(\pi\beta)^{-1/4}{\bar n}^{1/2}$ and
$\mu_D=-(\pi\beta)^{-1/4}{\bar
n}^{1/2}\,.$
The correct high degeneracy limit is recovered by using the (slightly modified) {\it
Gell-Mann Brueckner} approximations (including Hartree--Fock)
for the free energy and for
the chemical potential
\begin{equation}\label{pade5}
f_{GB}=-\frac{0.9163}{r_s}-0.08883
{\rm ln}\left[1+\frac{4.9262}{r_s^{0.7}}\right]\approx  -\frac{0.9163}{r_s}+0.0622
{\rm ln}r_s\,,  \end{equation}
\begin{equation}\label{pade6}
\mu_{GB}=-\frac{1.2217}{r_s}-0.08883
{\rm ln}\left[1+\frac{6.2208}{r_s^{0.7}}\right]\approx  -\frac{1.2217}{r_s}+0.0622
{\rm ln}r_s\,.
\end{equation}
The free energy is now  equal to
the {\it internal energy at $T=0$} and reads, according to Carr
and Maradudin,
\[
\frac{U}{N}=\frac{2.21}{r_s^2}-\frac{0.916}{r_s}+0.0622{\rm ln}r_s  -
0.096+0.018r_s{\rm ln}r_s+\cdots\,.
\]
The Brueckner parameter $r_s$ is given by
$r_s^3=3/(4\pi n)$. While the Hartree Fock term, i.e., the $1/r_s$ term
in  (\ref{pade5},\ref{pade6}), was retained unaffected, the  additional terms in
these equations and in formulae  (\ref{pade1},\ref{pade2}) were modified, or
{\it fitted},  respectively, such that (\ref{pade1},\ref{pade2}) meet the
numerical data {\it in between}, where the analytical limiting  formulae are not
applicable.
\begin{figure}
\centerline{\psfig{figure=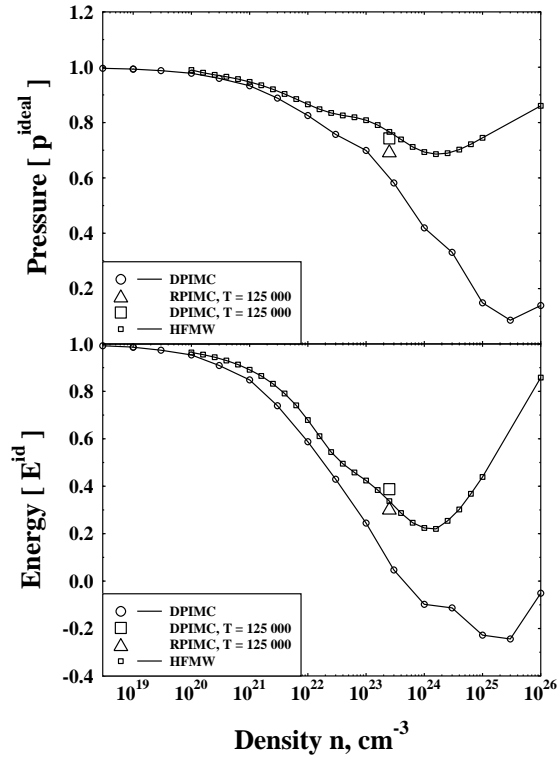,height=10cm}}
\caption{
Energy and pressure isotherms for $100,000 K$ (solid line with circles).
Solid line with small squares --
Hartree Fock (HF)  and Montroll--Ward (MW) approximation.
Reference data: triangle -- RPIMC (see Militzer et al.
\cite{militzer-etal.00}), large square -- DPIMC.}
\end{figure}

Consider now the proton contributions.
In the low density  regime, the proton formulae are practically
the same as for the  electrons, whereas in the high density limit we adjust the
formulae to (classical) Monte--Carlo (MC) data.    We have for the free energy
density and for the chemical potential
\begin{equation}\label{pade7}  -
\frac{f_p}{k_BTn_p}=\frac{(-f_p^{\rm int}/k_BTn_p)_D  [1-a{\tilde
n}_p^{2/3}(f_p^{\rm int}/k_BTn_p)_{MC}]}  {1-a{\tilde n}_p^{1/2}[{\tilde
n}_p^{1/2}/(\frac{f_p^{\rm int}}{k_BTn_P})_D  +{\tilde n}_p^{1/6}(\frac{f_p^{\rm
int}}{k_BTn_p    })_D]}\,  \end{equation}
\begin{equation}\label{pade8}
-\frac{\mu_p}{k_BT}=\frac{(-
\mu_p^{\rm int}/k_BT)_D  [1-2a{\tilde n}_p^{2/3}(\mu_p^{\rm int}/k_BTn_p)_{MC}]}
{1-2a{\tilde n}_p^{1/2}[{\tilde n}_p^{1/2}/(\frac{\mu_p^{\rm
int}}{k_BT})_D+{\tilde n}_p^{1/6}(\frac{\mu_p^{\rm int}}{k_BT})_D]}\,.
\end{equation}
We used the abbreviation  $a$ which depends only on temperature
\begin{equation}\label{pade9}
a=\sqrt{\pi^3k_BT}\Big\{\frac{1}{2}\Big[1+\sqrt{\frac{k_BT}{4\pi}}
{\rm exp}\Big(\frac{\sqrt{\pi}/2}{{\rm ln}{(4/k_BT)^{1/6}}-2\sqrt{k_BT}}\Big)\Big]  -
0.29931\Big\}\,.  \end{equation}
For the protons, we introduce the
dimensionless density to be used in the plasma
parameter $\Gamma$, namely
${\tilde
n}_p=\frac{8}{(k_BT)^3}n_p$, $\Gamma=\Big(\frac{4}{3}\pi  {\tilde
n}_p\Big)^{1/3}\,$.
The Debye approximations (i.e., the low
density case) for free energy  density and chemical potential read
$(-f_p^{\rm
int}/k_BTn_p)_D=2.1605{\tilde n}_p^{1/2}\,$
and
$(-\mu_p^{\rm
int}/k_BT)_D=\frac{3}{2}2.1605{\tilde n}_p^{1/2}\,$.
In the high density region, we use a fit to  (classical OCP) Monte--Carlo data.

For the free energy we write
\begin{eqnarray}\label{pade12a}
&&(f_p^{\rm int}/k_BTn_p)_{MC}=  -0.8946\Gamma +3.266\Gamma^{1/4}-0.5012
\ln\Gamma-2.809\nonumber\\  &&-\frac{r_s{\tilde
n}_p^{1/3}}{1+r_s^2}\Big[0.0933+1.0941{\tilde n}_p^{-1/4}-0.343{\tilde n}_p^{-
1/3}\Big]\,,  \end{eqnarray}  and for the chemical potential
\begin{eqnarray}\label{pade13}  &&(\mu_p^{\rm int}/k_BT)_{MC}  =-
1.1928\Gamma+3.5382\Gamma^{1/4}-0.5012\ln\Gamma-2.9761\nonumber\\  &&-
\frac{r_s{\tilde n}_p^{1/3}}{1+r_s^2}\Big[0.0933  +0.8206{\tilde n}_p^{-1/4}-
0.2287{\tilde n}_p^{-1/3}\Big]\,.  \end{eqnarray}  The {\it correlation part} of
the pressure for an $H$--plasma is then given by
\begin    {equation}\label{pade14}  p^{\rm corr}=p^{\rm corr}_e+p^{\rm
corr}_p=n_e\mu_e+n_p\mu_p-f_e-f_p\,.
\end{equation}
The contributions are
determined by  (\ref{pade1},\ref{pade2}) and (\ref{pade7},\ref{pade8}).  The
ideal pressure is given by  Fermi  integrals $I_{\nu}(\alpha)$  \[  p^{\rm
id}=k_BT\sum_a\frac{2s_a+1}{\Lambda_a^3}I_{3/2}(\alpha_a)\,\,,\,\,\,
\alpha_a=\mu_a/(k_BT)\,.
\]
The internal energy may be constructed from the excess free
energy  $f=\frac{F}{V}$ given above in addition to the ideal part according to
$U=F-T\frac{\partial F}{\partial T}\big|_{V={\rm const}}$\,,
where the ideal
free energy is given by
\[
F^{\rm id}=k_BTV\sum_a\frac{2s_a+1}{\Lambda_a^3}
\left\{\alpha_aI_{1/2}(\alpha_a)-I_{3/2}(\alpha_a)\right\}\,.
\]
At very high degeneracy,
the free energy is equal to the internal  energy.

\section{Hydrogen isotherms}

In this section we present results for the thermodynamic
functions of dense hydrogen versus density at constant
temperature. The PIMC simulations have been performed as
explained in  Filinov, V. S., et
al. \cite{filinov-etal.00jetpl},
and Filinov V. S., et al. \cite{ebeling-etal.00}.
Figures 1-3 show the simulation results together with the
Pad{\'e} results for
three hydrogen isotherms
$T=50,000$; $100,000$; and $125,000 K$. In all figures
the agreement between numerical and analytical
data is good for temperatures and densities, where
the coupling parameter $\Gamma $ is smaller than or equal to unity.
Reference points related to
RPIMC and DPIMC calculations in Fig. 1 correspond to data
available for the
temperature $T=62,500 K$.
\begin{figure}
\centerline{\psfig{figure=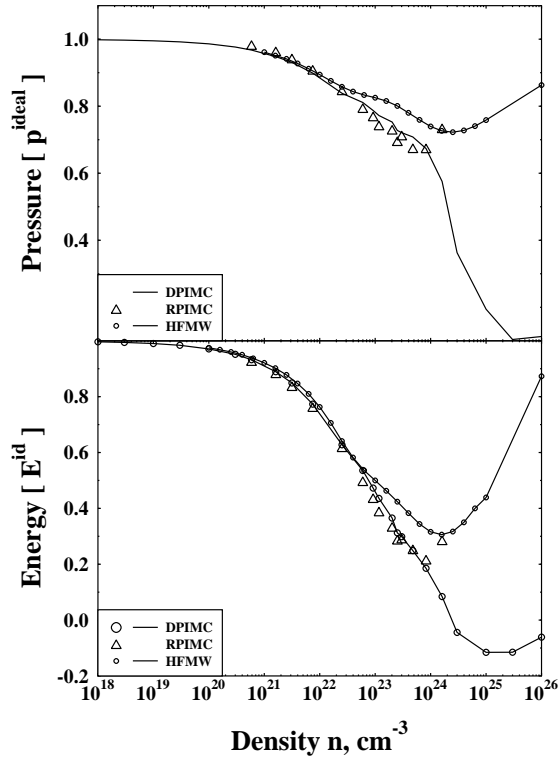,height=10cm}}
\caption{
Energy and pressure isotherms for $125,000 K$
(lower curves, solid line with larger or without circles).
Solid line with smaller circles -- Hartree Fock (HF)  and Montroll--Ward (MW) approximation
for $125 000 K $. Reference data: triangles   --  RPIMC
(see Militzer et al. \cite{militzer-etal.00}).}
\end{figure}

At low densities, pressure and energy are close to those of an
ideal plasma. Increasing the density above $10^{19}cm^{-3}$,
Coulomb interaction becomes important leading to a decrease of
pressure and energy.
Differences between analytical and
numerical calculations in Fig. 1  are observed
for densities above $10^{22}$
$cm^{-3}$ where the coupling parameter $\Gamma$ exceeds unity.
At temperatures of $100,000$ and $125,000 K$, differences are
observed for
$n$ above $5\times10^{22} cm^{-3}$. The degeneracy parameter $n_e \lambda ^3$
reaches here values of $0.4$.
At higher densities (around $10^{24}cm^{-3}$)
the degeneracy $n_e\lambda^3$ becomes larger than unity,
and the interaction parts of pressure and energy decrease as compared
to the respective ideal contributions,
which leads to an increase of pressure and energy.
At lower temperatures, this tendency is accompanied by the
vanishing of bound states, i.e., a transition from a
partially ionized plasma to a metal--like state.
This tendency is correctly reproduced by all methods, however the
density values of this increase vary.
In Fig. 3 we compare our results with data from RPIMC simulations
(see Militzer et al \cite{militzer-etal.00}). Obviously, the agreement is very
good up to densities below $10^{24}cm^{-3}$.
\section{Discussion}
This work is devoted to a Quantum Monte Carlo study of a
correlated proton-electron system  with degenerate electrons. We compared our
direct PIMC simulations with independent  restricted PIMC results of Militzer
and Ceperley and analytical formulae for isotherms corresponding to
$T=50,000$; $100,000$; and $125,000 K$. The values of $\Gamma $ and $n_e \Lambda_e^3 $
are varying in a wide range of values. This region is of
particular interest
as here pressure and temperature ionization occur and, therefore, an accurate
and consistent treatment of scattering and bound states  is crucial. We found
that the results
agree sufficiently well for coupling parameters smaller or equal to unity.
This is remarkable because analytical formulae, the DPIMC and RPIMC
simulations are completely independent and use essentially different
approximations.  We, therefore, expect that these results for hydrogen are
reliable  which is the main result of the present paper.  We hope that
our simulation results allow us to derive and test improved analytical
approximations in the future.

\section{Acknowledgements}  We acknowledge support by the Deutsche
Forschungsgemeinschaft (Sonderfor-\\
schungsbereich 198: M.B., D.K.,
and W.D.K;
Mercator-Programm:  V.S.F.). Our thanks are due to W. Ebeling and
M. Schlanges for stimulating discussions.
%

%
%

\end{document}